\documentclass[onecolumn,twoside,romanappendices,12pt]{IEEEtran}
\usepackage{setspace}
\usepackage{cite,color}%
\usepackage{amsfonts}%
\usepackage{amsmath,url}%
\usepackage{amssymb}%
\usepackage{graphicx}
\usepackage{balance}
\usepackage{amsmath,amssymb,graphicx,epsfig,url,cite,balance}
\begin{document}
\newcounter{eqncount}
\pagestyle{empty}
\thispagestyle{empty}
\title{On the Statistics of Cognitive Radio Capacity in Shadowing and Fast Fading Environments}
\author{\authorblockA{Muhammad Fainan Hanif~\IEEEauthorrefmark{1} and Peter J.
Smith~\IEEEauthorrefmark{1} \\
\authorblockA{\IEEEauthorrefmark{1}Department of Electrical and Computer Engineering, University of Canterbury,
Christchurch, New Zealand}\\
Email: mfh21@student.canterbury.ac.nz,~p.smith@elec.canterbury.ac.nz
\footnote{\hrule \vspace{0.8mm} \hspace{-3.5mm}Mailing address:
Dept. of Electrical and Computer Engineering, University of
Canterbury, Private Bag 4800, Christchurch, New Zealand. Phone:
+64-3-364-2987 Ext. 7157.
\newline Part of this work is to be presented at the IEEE CrownCom,
Germany, Hannover in 2009.\newline \texttt{Submitted to \emph{IEEE
Trans. Wireless Commun.}, June, 2009.}}}} \maketitle
\thispagestyle{empty} \pagestyle{empty}
\begin{abstract}
In this paper we consider the capacity of the cognitive radio (CR)
channel in different fading environments under a ``low interference
regime". First we derive the probability that the ``low interference
regime" holds under shadow fading as well as Rayleigh and Rician
fast fading conditions. We demonstrate that this is the dominant
case, especially in practical CR deployment scenarios. The capacity
of the CR channel depends critically on a power loss parameter,
$\alpha$, which governs how much transmit power the CR dedicates to
relaying the primary message. We derive a simple, accurate
approximation to $\alpha$ in Rayleigh and Rician fading environments
which gives considerable insight into system capacity. We also
investigate the effects of system parameters and propagation
environment on $\alpha$ and the CR capacity. In all cases, the use
of the approximation is shown to be extremely accurate.
\end{abstract}
\section{Introduction}
Until recently, the frequency bands below $3.5$ GHz were thought to
be severely congested. Due to the superior propagation conditions in
the lower frequencies there is a desire for all services to find a
place in this sought after ``real estate". However, spectrum
occupancy measurements performed in the United States \cite{report2}
show that spectrum scarcity cannot be confirmed by the measurements.
Instead, the apparent congestion is due to the way in which spectrum
is allocated into specific bands for specific services (i.e., fixed,
mobile and broadcasting) and then by the national regulatory
authorities who license the band/service combinations to private
owners. Therefore even when the licensed owner is not using their
spectrum, there is no access to other users, hence the apparent
congestion. In order to improve spectrum occupancy and utilization,
various regulatory bodies worldwide are considering the benefits
offered by cognitive radio (CR) \cite{report1}. The key idea behind
the deployment of CR is that greater utilization of spectrum can be
achieved if they are allowed to co-exist with the incumbent licensed
primary users (PUs) provided that they cause minimal interference.
The CRs must therefore learn from the radio environment and adapt
their parameters so that they can co-exist with the primary systems.
The CR field has proven to be a rich source of challenging problems.
A large number of papers have appeared on various aspects of CR,
namely spectrum sensing (see \cite{Ghasemi1,Tandra} and the
references therein), fundamental limits of spectrum sharing
\cite{Ghasemi}, information theoretic capacity limits
\cite{devroye,Maric,Jovicic,Jiang,Wu} etc.

The 2 user cognitive channel \cite{devroye,Maric,Jovicic,Jiang,Wu}
consists of a primary and a secondary user. It is very closely
related to the classic 2 user interference channel, see
\cite{Kramer} and references therein. The formulation of the CR
channel is due to Devroye \emph{et al.} \cite{devroye}. In this
channel, the CR has a non-causal knowledge of the intended message
of the primary and by employing dirty paper coding \cite{Costa} at
the CR transmitter it is able to circumvent the primary user's
interference to its receiver. However, the interference from the CR
to the primary receiver remains and has the potential to cause a
rate loss to the primary.

In recent work, Jovicic and Viswanath \cite{Jovicic} have studied
the fundamental limits of the capacity of the CR channel. They show
that if the CR is able to devote a part of its power to relaying the
primary message, it is possible to compensate for the rate loss to
the primary via this additional relay. They have provided exact
expressions for the PU and CR capacity of a 2 user CR channel when
the CR transmitter sustains a power loss by devoting a fraction,
$\alpha$, of its transmit power to relay the PU message.
Furthermore, they have provided an exact expression for $\alpha$
such that the PU rate remains the same as if there was no CR
interference. It should be stressed here that their system model is
such that at the expense of CR transmit power, the PU device is
always able to maintain a constant data rate. Hence, we focus on CR
rate, $\alpha$ and their statistics. They also assume that the PU
receiver uses a single user decoder. Their result holds for the so
called low interference regime when the interference-to-noise ratio
(INR) at the PU receiver is less than the signal-to-noise ratio
(SNR) at the CR receiver. The authors in \cite{Wu} also arrived at
the same results in their parallel but independent work.

The Jovicic and Viswanath study is for a static channel, i.e., the
direct and cross link gains are constants. In a system study, these
gains will be random and subject to distance dependent path loss and
shadow fading. Furthermore, the channel gains also experience fast
fading. As the channel gains are random variables, the power loss
parameter, $\alpha$, is also random.

In this paper we focus on the power loss, $\alpha$, the capacity of
the CR channel and the probability  that the ``low interference
regime'' holds. The motivation for this work arises from the fact
that maximum rate schemes for the CR in the low interference regime
\cite{Jovicic,Wu} and the achievable rate schemes for the high
interference regime \cite{Maric,Jiang} are very different. Hence, it
is of interest to identify which scenario is the most important. To
attack this question we propose a simple, physically based geometric
model for the CR, PU layout and compute the probability of the low
interference regime. Results are obviously limited to this
particular model but provide some insight into reasonable deployment
scenarios. Since the results show the low interference regime can be
dominant, it is also of interest to characterize CR performance via
the $\alpha$ parameter. In this area we make the following
contributions:
\begin{itemize}
\item Assuming lognormal shadowing, Rayleigh fading and path loss effects we derive the probability  that the ``low interference regime" holds.
We also extend the results to Rician fading channels.
\item In both Rayleigh and Rician fading environments we derive an approximation for $\alpha$ and its statistics.
This extremely accurate approximation leads to simple
interpretations of the effect of system parameters on the capacity.
\item Using the statistics of $\alpha$ we investigate the mean rate loss of the
CR and the cumulative distribution function (CDF) of the CR rates.
For both the above we show their dependence on the propagation
parameters.
\item We also show how the mean value of $\alpha$ varies with the CR
transmit power and therefore the CR coverage area.
\end{itemize}
This paper is organized as follows: Section~II describes the system
model. Section~III derives the probability  that the ``low
interference regime" holds and in Section~IV an approximation for
$\alpha$ is developed. Section~V presents analytical and simulation
results and some conclusions are given in Section~VI.
\section{System Model}
Consider a PU receiver in the center of a circular region of radius
$R_p$. The PU transmitter is located uniformly in an annulus of
outer radius $R_p$ and inner radius $R_{0}$ centered on the PU
receiver. It is to be noted that we place the PU receiver at the
center only for the sake of mathematical convenience (see Fig.
\ref{fig_1}). The use of the annulus restricts the length of the PU
link from becoming too small. This matches physical reality and also
avoids problems with the classical inverse power law relationship
between signal strength and distance \cite{Mai1}. In particular,
having a minimum distance, $R_0$, prevents the signal strength from
becoming infinite as the transmitter approaches the receiver.
Similarly, we assume that a CR transmitter is uniformly located in
the same annulus. Finally, a CR receiver is uniformly located in an
annulus centered on the CR transmitter. The dimensions of this
annulus are defined by an inner radius, $R_0$, and an outer radius,
$R_c$. This choice of system layout is asymmetric in the sense that
the PU receiver is at the center of its circular region whereas the
CR transmitter is at the center of its smaller region. This layout
is chosen for mathematical simplicity since the lengths of the CR-PU
and CR-CR links have a common simple distribution which leads to the
closed form analysis in Sec.~III. Following the work of Jovicic and
Viswanath \cite{Jovicic}, the four channel gains which define the
system are denoted $p, g, f, c$. In this paper, these complex
channel gains include shadow fading, path-loss and Rayleigh and
Rician fast fading effects. To introduce the required notation we
consider the link from the CR transmitter to the PU receiver, the CP
link. For this link we have:
\begin{equation}\label{linkdef}
|f|^2=\Gamma_{cp}|\tilde{f}|^2,
\end{equation}
where $|\tilde{f}|^2$ is an exponential random variable with unit
mean for Rayleigh channels or a noncentral $\chi^2$ variable for
Rician fading and $\Gamma_{cp}$ is the link gain. The link gain
comprises shadow fading and distance dependent path loss effects so
that,
\begin{equation}\label{signal}
\Gamma_{cp}=A_cL_{cp}r_{cp}^{-\gamma},
\end{equation}
where $A_c$ is a constant that depends on physical deployment
parameters such as antenna height, antenna gain, cable loss etc. In
(\ref{signal}) the variable $L_{cp}=10^{\tilde{X}_{cp}/10}$ is
lognormal, $\tilde{X}_{cp}$ is zero mean Gaussian and $r_{cp}$ is
the link distance. The standard deviation which defines the
lognormal is $\sigma$ (dB) and $\gamma$ is the path loss exponent.
For convenience, we also write $L_{cp}=e^{X_{cp}}$ so that
$X_{cp}=\beta \tilde{X}_{cp}$, $\beta=\ln(10)/10$ and
$\sigma_{sf}^2$ is the variance of $X_{cp}$. Hence, for the CP link
we have:
\begin{equation}\label{linkgain}
|f|^2=A_ce^{X_{cp}}r_{cp}^{-\gamma}|\tilde{f}|^2.
\end{equation}
The other three links are defined similarly where $\tilde{p},
\tilde{g}, \tilde{c}$ are standard exponentials for Rayleigh fading
and represent noncentral $\chi^2$ random variables for Rician
fading, $X_{pp}, X_{pc}, X_{cc},$ are Gaussians with the same
parameters as $X_{cp}$ and $r_{pp}, r_{pc}, r_{cc}$ are link
distances. However, for the links involving the PU transmitter we
assume a different constant $A_p$ in the model of link gains. The
parameters $A_p$ and $A_c$ are constants and all links are assumed
independent. The remaining parameters required are the transmit
powers of the PU and CR devices, given by $P_p$ and $P_c$
respectively, and the noise powers at the PU and CR receivers, given
by $N_p$ and $N_c$ respectively.

The physical model described above corresponds to the information
theoretic model shown in Fig.~\ref{fig_1n}. For fixed channel
coefficients, $p, g, f$ and $c$, Jovicic and Viswanath
\cite{Jovicic} compute the highest rate that the CR can achieve
subject to certain constraints using the model in Fig.~\ref{fig_1n}.
In this figure the arrow on the transmitter side indicates the
noncausal availability of the PU's message to the cognitive device
for dirty paper coding (DPC) purposes \cite{Costa}. A key constraint
is that the PU must not suffer any rate degradation due to the CR
and this is achieved by the CR dedicating a portion, $\alpha$, of
its transmit power to relaying the PU message. The parameter,
$\alpha$, is therefore central to determining the CR rate.
Furthermore, the results in \cite{Jovicic} are valid in the ``low
interference regime" defined by $a<1$ where:
\begin{equation}\label{defa}
a=\frac{\sqrt{N_c}\sqrt{\Gamma_{cp}}|\tilde{f}|}{\sqrt{N_p}\sqrt{\Gamma_{cc}}|\tilde{c}|}
=\frac{\sqrt{N_c}e^{X_{cp}/2}r_{cp}^{-\gamma/2}|\tilde{f}|}{\sqrt{N_p}e^{X_{cc}/2}r_{cc}^{-\gamma/2}|\tilde{c}|}.
\end{equation}
In this regime, the highest CR rate is given by
\begin{equation}\label{CRRate}
R_{CR}=\log_2\Bigg(1+\frac{|c|^2(1-\alpha)P_c}{N_c}\Bigg),
\end{equation}
with the power loss parameter, $\alpha$, defined by
\begin{equation}\label{alpha}
\alpha=\frac{|s|^2}{|t|^2}\Bigg[\frac{\sqrt{1+|t|^2(1+|s|^2)}-1}{1+|s|^2}\Bigg]^2,
\end{equation}
where $|s|=\sqrt{P_p}\sqrt{\Gamma_{pp}}|\tilde{p}|N_p^{-1/2}$ and
$|t|=\sqrt{P_c}\sqrt{\Gamma_{cp}}|\tilde{f}|N_p^{-1/2}$. Note that
the definitions of $\alpha$ and $R_c$ are conditional on $a<1$.
Since $a$ is a function of $\tilde{f}$ and $\tilde{c}$ we see that
both $\tilde{f}$ and $\tilde{c}$ are conditional random variables.
\section{The low interference regime}
Note that the $4$ paths which characterize the channels in
Figs.~\ref{fig_1} and \ref{fig_1n} can all be Rayleigh or Rician.
This leads to $16$ possible combinations of Rayleigh or Rician
channels. To make the study more concise we assume that the PP and
PC paths are Rayleigh and vary the CC and CP paths. Hence, we
consider the $4$ combinations where $\tilde{c}$ (CC) and $\tilde{f}$
(CP) can be Rician or Rayleigh. This is sensible since $\tilde{c}$,
$\tilde{f}$ affect both the low interference regime (\ref{defa}) and
the cognitive rate ($R_{CR}$ in (\ref{CRRate})), whereas the PP, PC
links only affect $R_{CR}$. The notation Ray/Rice etc. denotes the
nature of the $\tilde{f}/\tilde{c}$ variables or the CP/CC paths.
\subsection{Rayleigh/Rayleigh Scenario}
The low interference regime is defined by $a<1$, where $a$ is
defined in (\ref{defa}). The probability, $P(a<1)$, depends on the
distribution of $r_{cc}/r_{cp}$. Using standard transformation
theory \cite{pap}, some simple but lengthy calculations show that
the CDF of $r_{cc}/r_{cp}$ is given by (\ref{ratioofdis}). A sketch
proof is given in Appendix I.
\begin{figure*}[!t]
\normalsize \setcounter{eqncount}{\value{equation}}
\setcounter{equation}{6}
\begin{equation}\label{ratioofdis}
P\bigg(\frac{r_{cc}}{r_{cp}}<x\bigg) = \left\{ \begin{array}{ll} 0 &
\textrm{$x\leq\frac{R_0}{R_p}$}\\\\
\frac{0.5x^2(R_p^2-R_0^4x^{-4})-R_0^2(R_p^2-R_0^2x^{-2})}{(R_c^2-R_0^2)(R_p^2-R_0^2)}
&
\textrm{$\frac{R_0}{R_p}<x\leq\frac{R_c}{R_p}$}\\\\
\frac{0.5(R_c^4-R_0^4)-R_0^2(R_c^2-R_0^2)+(x^2R_p^2-R_c^2)(R_c^2-R_0^2)}{x^2(R_c^2-R_0^2)(R_p^2-R_0^2)} &
\textrm{$\frac{R_c}{R_p}<x\leq1$}\\\\
1-\frac{0.5R_c^4x^2+0.5R_0^4x^2-R_0^2R_c^2}{(R_c^2-R_0^2)(R_p^2-R_0^2)}
& \textrm{$1<x\leq\frac{R_c}{R_0}$}\\\\
1 & \textrm{$x>\frac{R_c}{R_0}$}
\end{array} \right.
\end{equation}
\setcounter{equation}{\value{eqncount}} \hrulefill
\end{figure*}
\addtocounter{equation}{1}
The CDF in (\ref{ratioofdis}) can be written as:
\begin{equation}\label{simpratioofdis}
P\bigg(\frac{r_{cc}}{r_{cp}}<x\bigg) = c_{i0}x^{-2}+c_{i1}+c_{i2}x^2
\quad \textrm{$i=1,2,3,4,5$}
\end{equation}
where $\Delta=(R_c^2-R_0^2)(R_p^2-R_0^2)$, $c_{10}=0$, $c_{11}=0$,
$c_{12}=0$, $c_{20}=0.5R_0^4/\Delta$, $c_{21}=-R_0^2R_p^2/\Delta$,
$c_{22}=0.5R_p^4/\Delta$, $c_{30}=0.5(R_0^4-R_c^4)/\Delta$,
$c_{31}=R_p^2(R_c^2-R_0^2)/\Delta$, $c_{32}=0$,
$c_{40}=-0.5R_c^4/\Delta$, $c_{41}=1+R_0^2R_c^2/\Delta$,
$c_{42}=-0.5R_0^4/\Delta$, $c_{50}=0$, $c_{51}=1$ and $c_{52}=0$.

Now $P(a<1)=P(a^2<1)$ can be written as $P(Y<Ke^XZ^{-\gamma})$ where
$Y={|\tilde{f}|^2}/{|\tilde{c}|^2}$, $K=N_p/N_c$, $X=X_{cc}-X_{cp}$
and $Z=r_{cc}/r_{cp}$. Thus the required probability is:
\setlength{\arraycolsep}{0.0em}
\begin{eqnarray}\label{lowintprob}
P(Y<Ke^XZ^{-\gamma})&{}={}&P(Z<K^{1/\gamma}e^{X/\gamma}Y^{-1/\gamma})\nonumber\\
&{}={}&E[P(Z<K^{1/\gamma}e^{X/\gamma}Y^{-1/\gamma}|X,Y)]\nonumber\\
&{}={}&E[P(Z<W|W)]\nonumber\\
&{}={}&\int_0^\infty P(Z<w)f_W(w)dw,
\end{eqnarray}
\setlength{\arraycolsep}{5pt}
\hspace{-1mm}where $W=K^{1/\gamma}e^{X/\gamma}Y^{-1/\gamma}$ and
$f_W(.)$ is the PDF of $W$. Note that $P(Z<w)$, given in
(\ref{simpratioofdis}), only contains constants and terms involving
$w^{\pm2}$. Hence, we need the following:
\begin{equation}\label{genint}
\int_\theta^\kappa\!\!w^{2m}f_W(w)dw=\int\!\!\int\!
(Ke^xy^{-1})^{2m/\gamma}f_{X,Y}(x,y)dxdy,
\end{equation}
where $m=-1,0,1$ and $f_{X,Y}(.)$ is the joint PDF of $X,Y$. Now,
since $W=K^{1/\gamma}e^{X/\gamma}Y^{-1/\gamma}$, the limits
$\theta\leq w\leq \kappa$ in (\ref{genint}) imply the following
limits for $x$:
\begin{displaymath}
\ln(\theta^{\gamma} K^{-1}y)\leq x\leq \ln(\kappa{^\gamma} K^{-1}y).
\end{displaymath}
Let $\ln(\theta^{\gamma} K^{-1}y)=A$ and $\ln(\kappa{^\gamma}
K^{-1}y)=B$, then noting that $f_{X,Y}(x,y)=f_X(x)f_Y(y)$, the
integral in (\ref{genint}) becomes:
\setlength{\arraycolsep}{0.0em}
\begin{eqnarray}\label{genint2}
\int_\theta^\kappa w^{2m}f_W(w)dw&{}={}&\int_0^\infty
K^{2m/\gamma}y^{-2m/\gamma}f_Y(y)\nonumber\\&&{\times}\:\int_{A}^{B}
e^{2mx/\gamma}f_X(x)dxdy.
\end{eqnarray}
\setlength{\arraycolsep}{5pt}
\hspace{-2mm}Since $X\sim\mathcal{N}(0,2\sigma_{sf}^2)$, the inner
integral in (\ref{genint2}) becomes:
\setlength{\arraycolsep}{0.0em}
\begin{eqnarray}\label{genint3}
\int_{A}^{B}
\!\!\!e^{2mx/\gamma}&{}f_X(x){}&dx=\exp\Bigg(\frac{4m^2\sigma_{sf}^2}{\gamma^2}\Bigg)\nonumber\\&&{\times}\:
\Bigg[\Phi\Bigg(\frac{B-\frac{4m\sigma_{sf}^2}{\gamma}}{\sqrt{2}\sigma_{sf}}\Bigg)-
\Phi\Bigg(\frac{A-\frac{4m\sigma_{sf}^2}{\gamma}}{\sqrt{2}\sigma_{sf}}\Bigg)\Bigg],\nonumber\\
\end{eqnarray}
\setlength{\arraycolsep}{5pt}
\hspace{-1.6mm}where $\Phi$ is the CDF of a standard Gaussian. Since
$f_Y(y)$ is the density function of the ratio of two standard
exponentials, it is given by \cite{Ghasemi}:
\begin{equation}\label{ratioexp}
f_Y(y)=\frac{1}{(1+y)^2}, \qquad y\geq 0
\end{equation}
Using (\ref{genint3}) and (\ref{ratioexp}), the total general
integral in (\ref{genint}) becomes:
\setlength{\arraycolsep}{0.0em}
\begin{eqnarray}\label{genintfinal}
\int_\theta^\kappa\!\!w^{2m}f_W(w)dw&{}={}&\int_0^\infty
\!\!K^{2m/\gamma}y^{-2m/\gamma}(1+y)^{-2}\exp\Bigg(\frac{4m^2\sigma_{sf}^2}{\gamma^2}\Bigg)\nonumber\\&&{\times}\:
\Bigg[\Phi\Bigg(\frac{B-\frac{4m\sigma_{sf}^2}{\gamma}}{\sqrt{2}\sigma_{sf}}\Bigg)-
\Phi\Bigg(\frac{A-\frac{4m\sigma_{sf}^2}{\gamma}}{\sqrt{2}\sigma_{sf}}\Bigg)\Bigg]dy\nonumber\\
&{}\triangleq{}&I(m,\theta,\kappa).
\end{eqnarray}
\setlength{\arraycolsep}{5pt}
Substituting (\ref{simpratioofdis}) and (\ref{genintfinal}) in
(\ref{lowintprob}) gives $P(a<1)$ as:
\setlength{\arraycolsep}{0.0em}
\begin{eqnarray}\label{lowintfinal}
P(a<1)&{}={}&P(Y<Ke^XZ^{-\gamma})\nonumber\\
&{}={}&\sum_{i=2}^5c_{i0}I(-1,\theta_i,\kappa_i)+c_{i1}I(0,\theta_i,\kappa_i)+c_{i2}I(1,\theta_i,\kappa_i)\nonumber\\
&{}={}&\sum_{i=2}^5\sum_{j=0}^2c_{ij}I(j-1,\theta_i,\kappa_i).
\end{eqnarray}
\setlength{\arraycolsep}{5pt}
\hspace{-1.0mm}Finally, it can be seen from the limits given in
(\ref{ratioofdis}) that $\kappa_i=\theta_{i+1}$. Hence, the final
expression for the probability of occurrence of the low interference
regime is:
\setlength{\arraycolsep}{0.0em}
\begin{eqnarray}\label{lowintfinal1}
P(a<1)&{}={}&\sum_{i=2}^5\sum_{j=0}^2c_{ij}I(j-1,\theta_i,\theta_{i+1}),
\end{eqnarray}
\setlength{\arraycolsep}{5pt}
\hspace{-1.5mm}where the $c_{ij}$ were defined after
(\ref{simpratioofdis}), $I(j-1,\theta_i,\theta_{i+1})$ is given in
(\ref{genintfinal}), $\theta_2=R_0/R_p$, $\theta_3=R_c/R_p$,
$\theta_4=1$, $\theta_5=R_c/R_0$ and $\theta_6=\infty$. Hence,
$P(a<1)$ can be derived in terms of a single numerical integral. For
numerical convenience, (\ref{genintfinal}) is rewritten using the
substitution $v=y(y+1)^{-1}$ so that a finite range integral over
$0<v<1$ is used for numerical results:
\setlength{\arraycolsep}{0.0em}
\begin{eqnarray}\label{genintfinalsim}
\int_\theta^\kappa\!\!\!w^{2m}&{}f_W(w)dw{}&\:=\int_0^1
\!\!\!K^{2m/\gamma}\Big(\frac{v}{1-v}\Big)^{-2m/\gamma}\exp\Bigg(\frac{4m^2\sigma_{sf}^2}{\gamma^2}\Bigg)\nonumber\\&&{\times}\:
\Bigg[\Phi\Bigg(\frac{B-\frac{4m\sigma_{sf}^2}{\gamma}}{\sqrt{2}\sigma_{sf}}\Bigg)-
\Phi\Bigg(\frac{A-\frac{4m\sigma_{sf}^2}{\gamma}}{\sqrt{2}\sigma_{sf}}\Bigg)\Bigg]dv\nonumber\\
&{}\triangleq{}&I(m,\theta,\kappa),
\end{eqnarray}
\setlength{\arraycolsep}{5pt}
\hspace{-1.68mm}where $\ln(\theta^{\gamma} K^{-1}\frac{v}{1-v})=A$
and $\ln(\kappa{^\gamma} K^{-1}\frac{v}{1-v})=B$. Further
simplification of $(\ref{genintfinal})$ appears difficult but the
result in (\ref{genintfinalsim}) is stable and rapid to compute.

It can be easily inferred from the above discussion that the
probability of low interference regime in (\ref{lowintfinal1})
depends on the ratio (\ref{ratioexp}) of random variables
representing fast fading in the interfering and direct links from
the point of view of the cognitive device. Hence, we focus on the
following three cases of interest as well.
\subsection{Rayleigh/Rician Scenario}
In this case the probability density function (PDF) of the ratio
$Y={|\tilde{f}|^2}/{|\tilde{c}|^2}$ is given by \cite{Himal}:
\begin{equation}\label{Himal1}
f_Y(y)=(K+1)\frac{y+(K+1)^2}{(y+K+1)^3}e^{-K+\frac{K^2+K}{y+K+1}},
\end{equation}
where $K$ is the Rician $K$ factor defined as the ratio of signal
power in the dominant component to the scattered power and $f_Y(y)$
represents the PDF of the ratio of a standard exponential to a
noncentral $\chi^2$ random variable. Now $P(a<1)$ can easily be
calculated by substituting (\ref{Himal1}) in (\ref{genint2}) and
evaluating (\ref{lowintfinal1}). However, as mentioned above the
substitution $v=y(y+1)^{-1}$ is again used to obtain the numerical
results.
\subsection{Rician/Rayleigh Scenario}
When the interfering signal is a Rician variable and the direct
signal follows Rayleigh distribution, the PDF of $Y$, after
correcting the expression in \cite{Himal}, is:
\begin{equation}\label{Himal2}
f_Y(y)=\frac{K(1+K)}{(y+Ky+1)^2}e^{-\frac{K}{y+Ky+1}}+\frac{1-K^2+y(1+2K+K^2)}{(y+Ky+1)^3}e^{-K+\frac{Ky+K^2y}{y+Ky+1}},
\end{equation}
where $K$ is the Rician $K$ factor defined as above.
\subsection{Rician/Rician Scenario}
In this final case, the PDF $f_Y(y)$ represents the ratio of two
noncentral $\chi^2$ variables. It is known that \cite{Kutz} this
ratio characterizes the \emph{doubly noncentral F-distribution}.
Assuming that the noncentral $\chi^2$ random variable in the
numerator of $Y$ has $\nu_1$ degrees of freedom, $\lambda_1$
non-centrality parameter and the noncentral $\chi^2$ variable in the
denominator has $\nu_2$ degrees of freedom and $\lambda_2$
non-centrality parameter, the PDF of $Y$ is given by \cite{Kutz}:
\setlength{\arraycolsep}{0.0em}
\begin{eqnarray}\label{F}
f_Y(y)&{}={}&\sum_{j=0}^\infty\sum_{k=0}^\infty\bigg[\frac{e^{-\lambda_1/2}(0.5\lambda_1)^j}{j!}\bigg]
\bigg[\frac{e^{-\lambda_2/2}(0.5\lambda_2)^k}{k!}\bigg]\bigg[B\big(0.5\nu_1+j,0.5\nu_2+k\big)\bigg]^{-1}\nonumber\\
&&\times\:y^{0.5\nu_1+j-1}(1+y)^{-0.5(\nu_1+\nu_2)-j-k},
\end{eqnarray}
\setlength{\arraycolsep}{5pt}
\hspace{-1.5mm} where $B(.,.)$ is the beta function. It is worth
mentioning that we use $\nu_1=\nu_2=2$ and $\lambda_1=\lambda_2=2K$
while employing the above PDF to evaluate the probabilities.
Although the doubly infinite sum in (\ref{F}) is undesirable,
satisfactory convergence was found with only $18$ terms. Hence, the
approach is rapid and stable computationally.
A comparison of simulated and analytical results is presented in
Figs.~\ref{fig2} and \ref{new_fig_1}. It can the seen that the
analytical formulae for all the cases shown perfectly agree with the
simulation results for different parameter values. A discussion of
these results is presented in Sec.~V.
\section{An Approximation For The Power Loss Parameter}
In this section we focus on the power loss parameter, $\alpha$,
which governs how much of the transmit power the CR dedicates to
relaying the primary message. The exact distribution of $\alpha$
appears to be rather complicated, even for fixed link gains (fixed
values of $\Gamma_{cp},\Gamma_{pc},\Gamma_{pp}$ and $\Gamma_{cc}$).
Hence, we consider an extremely simple approximation based on the
idea that $|s||t|$ is usually small and $|s||t|>>|t|$. This
approximation is motivated by the fact that the CP link is usually
very weak compared to the PP link. This stems from the common
scenario where the CRs will employ much lower transmit powers than
the PU as the CC paths are usually much shorter. With this
assumption it follows that $|t|^2(1+|s|^2)$ is small and we have:
\setlength{\arraycolsep}{0.0em}
\begin{eqnarray}\label{alphasimplify}
\sqrt{\alpha}&{}={}&\frac{|s|}{|t|}\Bigg[\frac{\big(1+|t|^2(1+|s|^2)\big)^{1/2}-1}{1+|s|^2}\Bigg]\nonumber\\
&{}\approx{}&\frac{|s|}{|t|}\Bigg[\frac{1/2|t|^2(1+|s|^2)}{1+|s|^2}\Bigg]\nonumber\\
&{}={}&\frac{|s||t|}{2}\nonumber\\
&{}={}&\sqrt{\alpha_{approx}}.
\end{eqnarray}
\setlength{\arraycolsep}{5pt}
\noindent Expanding $\alpha_{approx}$ we have:
\begin{equation}\label{alphaapprox}
\alpha_{approx}= \frac{A_pA_cP_p P_c}{4
N_p^2}e^{(X_{pp}+X_{cp})}r_{pp}^{-\gamma}r_{cp}^{-\gamma}|{\tilde{p}}|^2|{\tilde{f}}|^2.
\end{equation}

This approximation is very effective for low values of
$\alpha_{approx}$, but is poor for larger values since
$\alpha_{approx}$ is unbounded whereas $0<\alpha<1$. To improve the
approximation, we use the conditional distribution of
$\alpha_{approx}$ given that $\alpha_{approx}<1$. This conditional
variable is denoted, ${\hat{\alpha}}$. The exact distribution of
${\hat{\alpha}}$ is difficult for variable link gains. However, the
approximation has a simple representation which leads to
considerable insight into the power loss and how it relates to
system parameters. For example $\alpha_{approx}$ is proportional to
$|s|^2|t|^2$ so that high power loss may be caused by high values of
$|s|$ or $|t|$ or moderate values of both. Now $|s|$ and $|t|$
relate to the PP and CP links respectively. Hence, the CR is forced
to use high power relaying the PU message when the CP link is
strong. This is obvious as the relay action needs to make up for the
strong interference caused by the CR. The second scenario is that
the CR has high $\alpha$ when the PP link is strong. This is less
obvious, but here the PU rate is high and a substantial relaying
effort is required to counteract the efforts of interference on a
high rate link. This is discussed further in Section~V. It is worth
noting that the condition $|s||t|>>|t|$ holds good only for some
specific values of channel parameters which support the assumption
that the CP link is usually much weaker than the PP link. Hence,
although it is motivated by a sensible physical scenario, it
requires verification. Results in Figs.~\ref{fig3}, \ref{fig6} and
\ref{fig4} show that it works very well. For fixed link gains, the
distribution of ${\hat{\alpha}}$ is:
\setlength{\arraycolsep}{0.0em}
\begin{eqnarray}\label{begin}
P(\alpha_{approx}<x|\alpha_{approx}<1)&{}={}&P(\hat{\alpha}<x)\nonumber\\
&{}={}&\frac{P(\alpha_{approx}<x)}{P(\alpha_{approx}<1)}.
\end{eqnarray}
\setlength{\arraycolsep}{5pt}
\noindent Thus, to compute the distribution function of
$\hat{\alpha}$ we need to determine $P(\alpha_{approx}<x)$ which can
be written as
\begin{equation}\label{firststep}
P(\alpha_{approx}<x)=P(|s|^2|t|^2<4x).
\end{equation}
Let $E(|s|^2)=\mu_s$, $E(|t|^2)=\mu_t$ with
$\mu_s=P_p\Gamma_{pp}/N_p$ and $\mu_t=P_c\Gamma_{cp}/N_p$. Further,
suppose that $U$, $V$ and $W$ are defined by $U=|\tilde{f}|^2$,
$V=|\tilde{c}|^2$ and $W=|\tilde{p}|^2$. We wish to derive
$P\big(WU<\frac{4x}{\mu_s\mu_t}\big)$, i.e., (\ref{firststep}),
subject to the condition $a<1$, which implies that $U<V/d$, where
$d=(N_c/N_p)(\Gamma_{cp}/\Gamma_{cc})$. Assuming $\zeta=4/\mu_s
\mu_t$ the required conditional CDF is given by:
\setlength{\arraycolsep}{0.0em}
\begin{eqnarray}\label{condDis}
&{}P{}&\bigg(UW<\zeta
x|U<\frac{V}{d}\bigg)\nonumber\\&{}={}&\frac{P\bigg(U\leq
\frac{\zeta x}{W},U<\frac{V}{d}\bigg)}{P\big(U<\frac{V}{d}\big)}\nonumber\\
&{}={}&\frac{\int_w\int_v P(U<\min(\frac{\zeta x}{w},\frac{v}{d}))f_W(w)f_V(v)dvdw}{\int_0^\infty P(U<\frac{v}{d})f_V(v)dv}\nonumber\\
&{}={}&\frac{\int_{w=0}^\infty\int_{v=0}^{\zeta xd/w}
P(U<\frac{v}{d})f_W(w)f_V(v)dvdw+\int_{w=0}^\infty\int_{v=\zeta
xd/w}^{\infty}
P(U<\frac{\zeta x}{w})f_W(w)f_V(v)dvdw}{\int_0^\infty P(U<\frac{v}{d})f_V(v)dv}\nonumber\\
&{}={}&\frac{\int_{v=0}^\infty\int_{w=0}^{\zeta xd/v}
P(U<\frac{v}{d})f_V(v)f_W(w)dwdv+\int_{w=0}^\infty\int_{v=\zeta
xd/w}^{\infty}P(U<\frac{\zeta x}{w})f_W(w)f_V(v)dvdw}{\int_0^\infty
P(U<\frac{v}{d})f_V(v)dv}\nonumber\\
&{}={}&\frac{\int_{v=0}^\infty F_W(\zeta
xd/v)F_U(v/d)f_V(v)dv+\int_{w=0}^\infty
F_U(\zeta x/w)(1-F_V(\zeta xd/w))f_W(w)dw}{\int_0^\infty F_U(v/d)f_V(v)dv}.\nonumber\\
\end{eqnarray}
\setlength{\arraycolsep}{5pt}
In the above derivation $f_U(u)$ and $F_U(u)$ represent the PDF and
CDF of $U$ respectively with similar definitions for $V$ and $W$.
With the general result in (\ref{condDis}), the CDF of
$\alpha_{approx}$ can be determined for any fading combinations
across the links of the CR interference channel. In most cases where
Rician fading occurs (\ref{condDis}) has to be computed via infinite
series expansions or numerical integration. In the Rayleigh fading
scenario a closed form solution is possible. Since for this case all
the distribution and density functions given in (\ref{condDis}) are
those of a standard unit mean exponential random variable, after a
few algebraic manipulations (details given in Appendix~II) and the
substitution $\zeta=4/\mu_s \mu_t$ we have:
\setlength{\arraycolsep}{0.0em}
\begin{eqnarray}\label{approxcdf}
P(\alpha_{approx}<x)&{}={}&1-\sqrt{\frac{16(1+d)x}{\mu_s\mu_t}}K_1\bigg(\sqrt{\frac{16(1+d)x}{\mu_s\mu_t}}\bigg),
\end{eqnarray}
\setlength{\arraycolsep}{5pt}
\hspace{-1.5mm}where $K_1(.)$ represents the modified Bessel
function of the second kind. Using the expression given in
(\ref{approxcdf}), the CDF of $\hat{\alpha}$ follows from
(\ref{begin}). Note that the CDF of $R_{CR}$ in (\ref{CRRate}) can
easily be obtained in the form of a single numerical integral for
fixed link gains as below:
\setlength{\arraycolsep}{0.0em}
\begin{eqnarray}\label{cdfRCR}
P(R_{CR}<x)&{}={}&P\bigg(|c|^2(1-\alpha)<(2^x-1)\frac{N_c}{P_c}\bigg)\nonumber\\
&{}={}&E\bigg[P\bigg(\alpha>1-\frac{(2^x-1)N_c}{|c|^2P_c}\bigg)\bigg]\nonumber\\
&{}={}&\int_0^\infty\bigg(1-F_\alpha\bigg(1-\frac{(2^x-1)N_c}{|c|^2P_c}\bigg)\bigg)f_c(c)dc
\end{eqnarray}
\setlength{\arraycolsep}{5pt}
where $F_\alpha(.)$ is the CDF of $\alpha$ in (\ref{approxcdf}) and
$f_c(c)$ is the PDF of $c$.
\section{Results}
In the results section, the default parameters are $\sigma=8$ dB,
$\gamma=3.5$, $R_0=1$ m, $R_c=100$ m, $R_p=1000$ m and
$N_p=N_c=P_p=P_c=1$. The parameter $A_p$ is determined by ensuring
that the PP link has an SNR $\geq5$ dB 95\% of the time in the
absence of any interference. Similarly, assuming that both PU and CR
devices have same threshold power at their cell edges, the constant
$A_c=A_p(R_p/R_c)^{-\gamma}$. Unless otherwise stated these
parameters are used in the following.
\subsection{Low interference regime}
In Figs.~\ref{fig2} and \ref{new_fig_1} we show that the low
interference regime, $a<1$, is the dominant scenario when the CR
coverage area is small compared to that of PU. For typical values of
$\gamma\in[3,4]$ and $\sigma\in[6,12]$ dB we find that $P(a<1)$ is
usually well over 90\% when $R_c$ is less than 20\% of $R_p$. As
expected, when $R_c$ approaches $R_p$ the probability drops and
reaches $P(a<1)=0.5$ when $R_c=R_p$. Note that this is only the case
when all the channel parameters are the same for the CC and CP
links. From Fig.~\ref{new_fig_1} we observe that the results are
reasonably insensitive to the type of fast fading. This is due to
the lesser importance of the fast fading compared to the large
effects of shadowing and path loss. Figure~\ref{fig2} also verifies
the analytical result in (\ref{lowintfinal}).

The relationship between $P(a<1)$ and the system parameters is
easily seen from (\ref{defa}) which contains the term
$\big(r_{cc}/r_{cp}\big)^{\gamma/2}\exp\big((X_{cp}-X_{cc})/2\big)$.
When $R_c<<R_p$, this term decreases dramatically as $\gamma$
increases (i.e., $P(a<1)$ increases) and as $\sigma$ increases the
term increases (hence $P(a<1)$ decreases). Also, as $R_c$ increases
$r_{cc}/r_{cp}$ tends to increase which in turn decreases $P(a<1)$.
When $R_c\approx R_p$ the low and high interference scenarios occur
with similar frequency (Fig.~\ref{new_fig_1}). This may be a
relevant system consideration if CRs were to be introduced in
cellular bands where the cellular hot spots, indoor micro-cells and
CRs will have roughly the same coverage radius. Note that $a$ is
independent of the transmit power, $P_c$. These conclusions are all
verified in Figs.~\ref{fig2} and \ref{new_fig_1}.
\subsection{Statistics of the power loss parameter, $\alpha$}
Figures~\ref{fig3}-\ref{fig6} all focus on the properties of
$\alpha$. Figure \ref{fig3} shows that the probability density
function (PDF) of $\alpha$ is extremely well approximated by the PDF
of $\hat{\alpha}$ in both Rayleigh and Rician fading channels. In
Fig.~\ref{fig5} we see that $E(\alpha)$ increases with increasing
values of $R_c/R_p$ and decreasing values of $\gamma$. This can be
seen from (\ref{alphaapprox}) where $\alpha_{approx}$ contains a
$(r_{pp}r_{cp})^{-\gamma}$ term which increases as $\gamma$
decreases, thus increasing the mean value of $\alpha$. The increase
of $E(\alpha)$ with $R_c$ follows from the corresponding increase in
$P_c$ to cater for larger $R_c$ values. Increasing the line of sight
(LOS) factor tends to increase $E(\alpha)$ although the effect is
minor compared to changes in $\gamma$, $\sigma$ or $R_c/R_p$. In
Fig.~\ref{fig5} we have limited $R_c/R_p$ to a maximum of $30\%$ as
beyond this value the high interference regime is also present with
a non-negligible probability. In Fig.~\ref{fig6} we see the
analytical CDF in (\ref{approxcdf}) verified by simulations for five
different scenarios of fixed link gains (simply the first five
simulated values of $\Gamma_{pp}$ and $\Gamma_{cp}$). Note that in
the different curves each correspond to a random drop of the PU and
CR transmitters. This fixes the distance and shadow fading terms in
the link gains in (\ref{signal}), thereby the remaining variation in
(\ref{linkdef}) is only Rayleigh. By computing a large number of
such CDFs and averaging them over the link gains a single CDF can be
constructed. This approach can be used to find the PDF of
$\hat{\alpha}$ as shown in Fig.~\ref{fig3}. Note that the curves in
Fig.~\ref{fig6} do not match exactly since the analysis is for
$\hat{\alpha}$ and the simulation is for $\alpha$.
\subsection{CR rates}
Figures~\ref{fig4}-\ref{fig8} focus on the CR rate $R_{CR}$.
Figure~\ref{fig4} demonstrates that the use of $\hat{\alpha}$ is not
only accurate for $\alpha$ but also leads to excellent agreement for
the CR rate, $R_{CR}$. This agreement holds over the whole range and
for all typical parameter values. Figure \ref{fig7} shows the \%
loss given by
$[R_{CR}(\alpha=0)-R_{CR}(\alpha)]/[R_{CR}(\alpha=0)]\%$. The loss
decreases as $\gamma$ increases, as discussed above, and increases
with $\sigma$. From (\ref{alphaapprox}) it is clear that increasing
$\sigma$ lends to larger values of $\exp(X_{pp}+X_{cp})$ which in
turn increases $\alpha$ and the rate loss. Note that the rate loss
is minor for $\sigma\in[8-10]$ dB with $R_c=R_p/10$. In a companion
paper \cite{ICC}, we show that the interference to the PU increases
with $\sigma$ and decreases with $\gamma$. These results reinforce
this observation, i.e., when the PU suffers more interference
($\sigma$ is larger) the CR has to devote a higher part of its power
to the PU. Consequently the percentage rate loss is higher. Again
the effect of $K$, the LOS factor, is minor compared to $\gamma$ and
$\sigma$.

Finally, in Fig.~\ref{fig8} we investigate the gains available to
the CR through increasing transmit power. The original transmit
power, $P_c$, is scaled by $\beta$ and the mean CR rate is simulated
over a range of $\beta$ values. Due to the relaying performed by the
CR, the PU rate is unaffected by the CR for any values of $\beta$
and so the CR is able to boost its own rate with higher transmit
power. Clearly the increased value of $\alpha$ for higher values of
$\beta$ is outweighed by the larger $P_c$ value and so the CR does
achieve an overall rate gain. In a very coarse way these results
suggest that multiple CRs may be able to co-exist with the PU since
the increased interference power might be due to several CRs and the
rate gain might be spread over several CRs. Of course, this
conclusion is speculative as the analysis is only valid for a single
CR.
\section{Conclusion}
In this paper we derive the probability that the ``low interference
regime'' holds and demonstrate the conditions under which this is
the dominant scenario. We show that the probability of the low
interference regime is significantly influenced by the system
geometry. When the CR coverage radius is small relative to the PU
radius, the low interference regime is dominant. On the other hand,
when the CR coverage radius approaches a value similar to the PU
coverage radius, the low and high interference regimes both occur
with roughly equal probability. In addition, we have derived a
simple, accurate approximation to $\alpha$ which gives considerable
insight into the system capacity. The $\alpha$ approximation shows
that the mean value of $\alpha$ is increased by small values of
$\gamma$, large CR coverage zones and higher $\sigma$ values. This
in turn decreases CR rates due to small values of $\gamma$, large CR
coverage zones and $\sigma$. The effect of the LOS strength is shown
to be minor and all results appear to be insensitive to the type of
fast fading. Finally, we have shown that the CR can increase its own
rate with higher transmit powers, although the relationship is only
slowly increasing as expected.
\appendices
\section{}
The variable $r_{cc}$ represents the distance of the CR link where
the receiver is uniformly located in an annulus of dimension
$[R_0,R_c]$ around the transmitter. Similarly, $r_{cp}$ describes
the the distance of the CR transmitter to the PU receiver where the
CR transmitter is uniformly located in an annulus of dimension
$[R_0,R_p]$ around the PU receiver. To evaluate the distribution of
$r_{cc}/r_{cp}$, we proceed as:
\setlength{\arraycolsep}{0.0em}
\begin{eqnarray}\label{app1}
P(r_{cc}<xr_{cp})&{}={}&E_{r_{cp}}[P(r_{cc}<xr_{cp}|r_{cp})]\nonumber\\
&{}={}&\int_{\alpha}^{\beta}\frac{2r_{cp}(x^2r_{cp}^2-R_0^2)}{(R_c^2-R_0^2)(R_p^2-R_0^2)}\:dr_{cp}\nonumber\\
&{}={}&\frac{0.5x^2(\beta^4-\alpha^4)-R_0^2(\beta^2-\alpha^2)}{(R_c^2-R_0^2)(R_p^2-R_0^2)},
\end{eqnarray}
\setlength{\arraycolsep}{5pt}
where we have used the facts that the PDF of the variable $r_{cp}$
is given by $2r_{cp}/(R_p^2-R_0^2)$ and that
$P(r_{cc}<xr_{cp})=(x^2r_{cp}^2-R_0^2)/(R_c^2-R_0^2)$. A little
inspection reveals that the random variable $r_{cp}$ takes on the
values $\alpha<x\leq \beta$ corresponding to the three different
ranges of $x$ as below:
\begin{itemize}
\item for $R_0/R_p<x<R_c/R_p$, $r_{cp}$ ranges from $\alpha=R_0/x$ to
$\beta=R_p$,
\item for $R_c/R_p<x<1$, $r_{cp}$ has a range from $\alpha=R_0/x$ to
$\beta=R_c/x$, and
\item for $1<x<R_c/R_0$, $r_{cp}$ spans a range from $\alpha=R_0$ to $\beta=R_c/x$.
\end{itemize}
Hence, using the above ranges of $x$ and $r_{cp}$ in (\ref{app1}),
some mathematical manipulations lead to (\ref{ratioofdis}).
\section{}
When there is Rayleigh fading in all links of the CR interference
channel, the distribution and density functions given in
(\ref{condDis}) are those of a standard unit mean exponential random
variable. Thus, with this substitution in (\ref{condDis}) we get:
\setlength{\arraycolsep}{0.0em}
\begin{eqnarray}\label{appII}
P\bigg(UW<\zeta x|U<\frac{V}{d}\bigg)&{}={}&\frac{\int_0^\infty
(1-e^{-\zeta xd/v})(1-e^{-v/d})e^{-v}dv+\int_0^\infty(1-e^{-\zeta
x/w})e^{-w}e^{-\zeta xd/w}dw}{\int_0^\infty
(1-e^{-v/d})e^{-v}dv}\nonumber\\
&{}={}&1+\frac{\int_0^\infty e^{-\zeta
xd/v-v(1+1/d)}dv-\int_0^\infty
e^{-w-\zeta/w(x+xd)}dw}{1-d/(1+d)}\nonumber\\
&{}={}&1+(d+1)\bigg[\int_0^\infty e^{-\zeta
xd/v-v(1+d)/d}dv-\int_0^\infty
e^{-w-\zeta x(1+d)/w}dw\bigg]\nonumber\\
&{}\stackrel{a}{=}{}&1+(d+1)\bigg[\int_0^\infty
e^{-\zeta xd/v-v(1+d)/d}dv-(1+d)/d\int_0^\infty e^{-v(1+d)/d-\zeta xd/v}dv\bigg]\nonumber\\
&{}={}& 1-(d+1)/d\int_0^\infty e^{-\zeta xd/v-v(1+d)/d}dv\nonumber\\
&{}\stackrel{b}{=}{}&1-\int_0^\infty e^{-\zeta x(1+d)/t-t}dt.
\end{eqnarray}
\setlength{\arraycolsep}{5pt}
where in both $a$ and $b$ above we have used the substitutions
$w=v(1+d)/d$ and $t=v(1+d)/d$ respectively. Now using $\zeta=4/\mu_s
\mu_t$ and evaluating the integral in the last equality using a
standard result in \cite{int} we arrive at (\ref{approxcdf}).
\balance
%\nocite{*}
\bibliographystyle{IEEEtran}
\bibliography{IEEEabrv,jovicic}
\newpage
\begin{figure}[t]
\centering
\includegraphics[width=0.55\columnwidth]{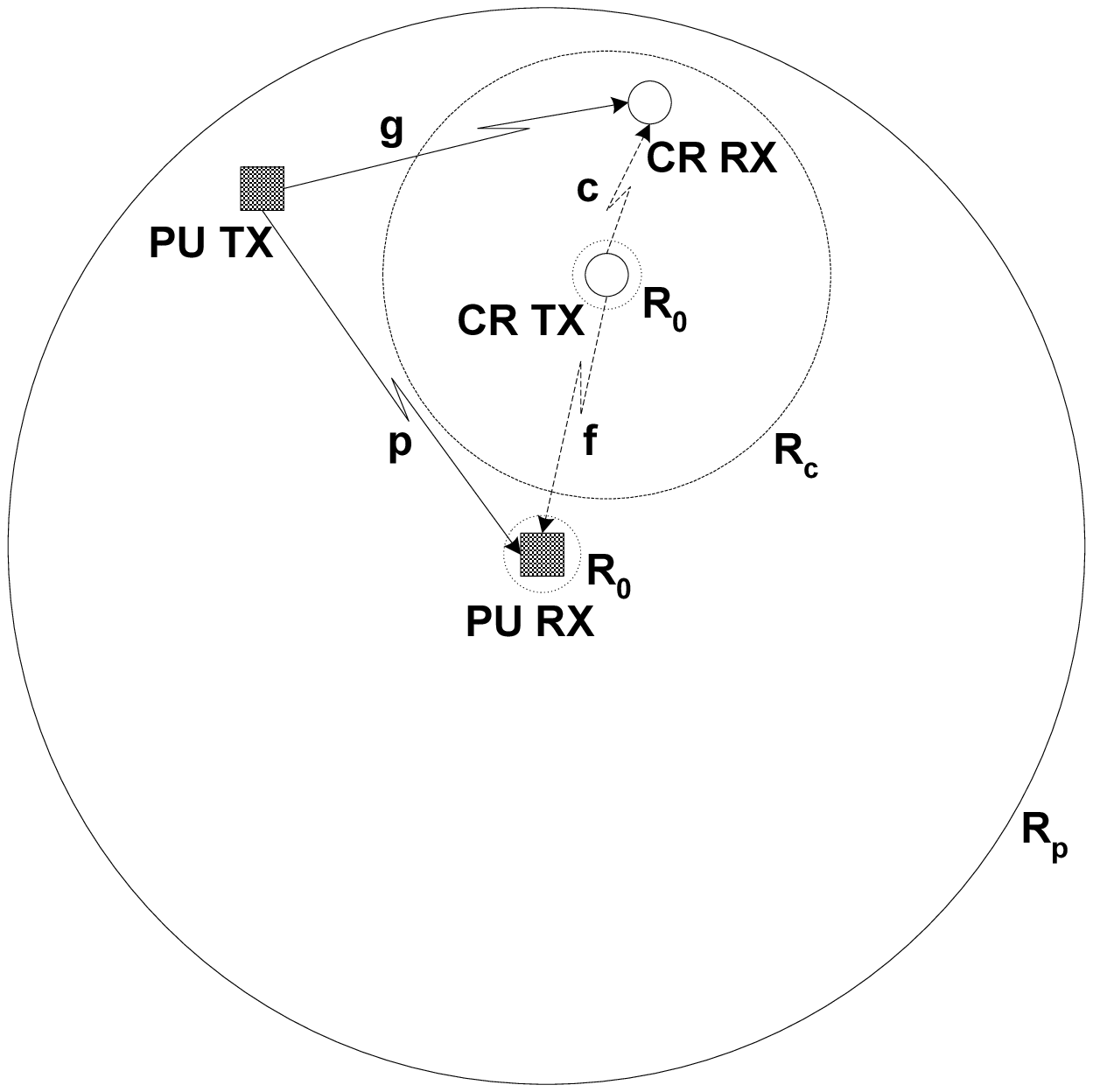}
\caption{System model.} \label{fig_1}
\end{figure}
\begin{figure}[t]
\centering
\includegraphics[width=0.65\columnwidth]{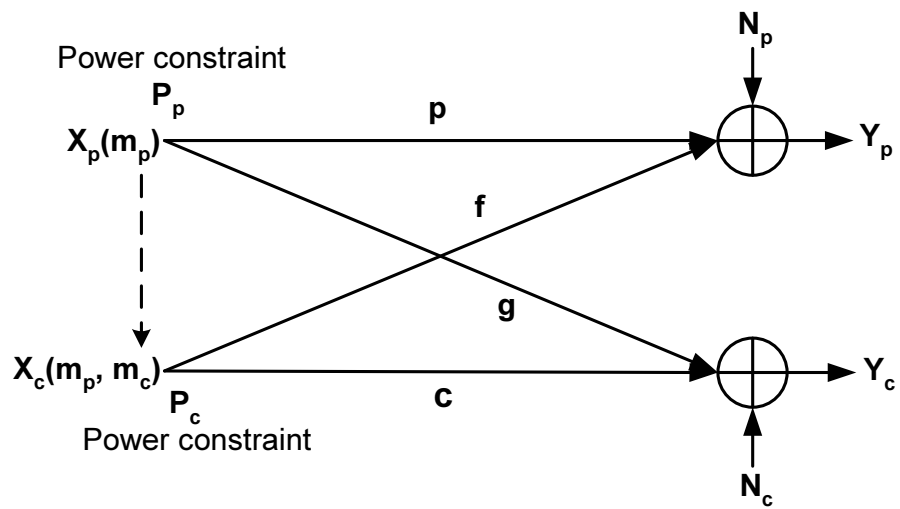}
\caption{Information theoretic model (taken from \cite{Jovicic}).}
\label{fig_1n}
\end{figure}
\begin{figure}[t]
\centering
\includegraphics[width=0.75\columnwidth]{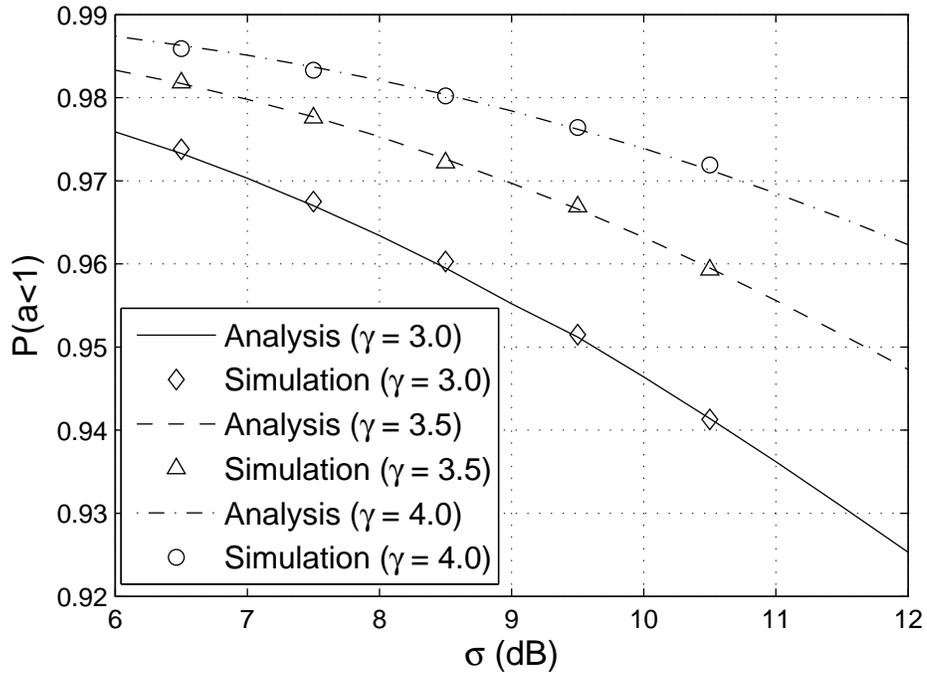}
\caption{Probability of occurrence of the low interference regime as
a function of shadow fading variance, $\sigma$ (dB) for Ray/Ray
scenario.} \label{fig2}
\end{figure}
\begin{figure}[t]
\centering
\includegraphics[width=0.75\columnwidth]{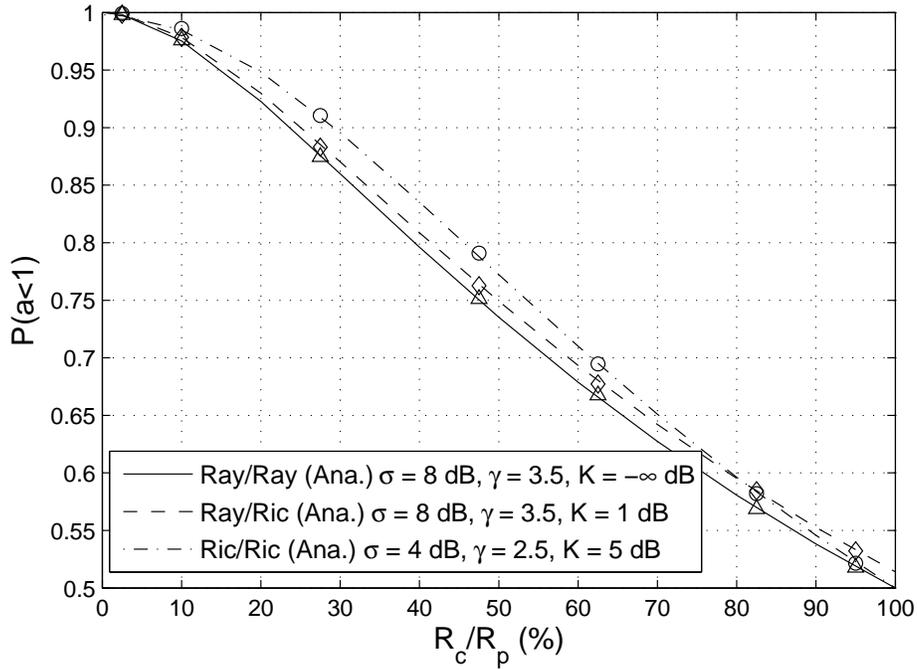}
\caption{Probability of occurrence of the low interference regime as
a function of the ratio $R_c/R_p$ for different fading scenarios.
Simulation values are shown by markers on the analytical
curves.}\label{new_fig_1}
\end{figure}
\begin{figure}[t]
\centering
\includegraphics[width=0.75\columnwidth]{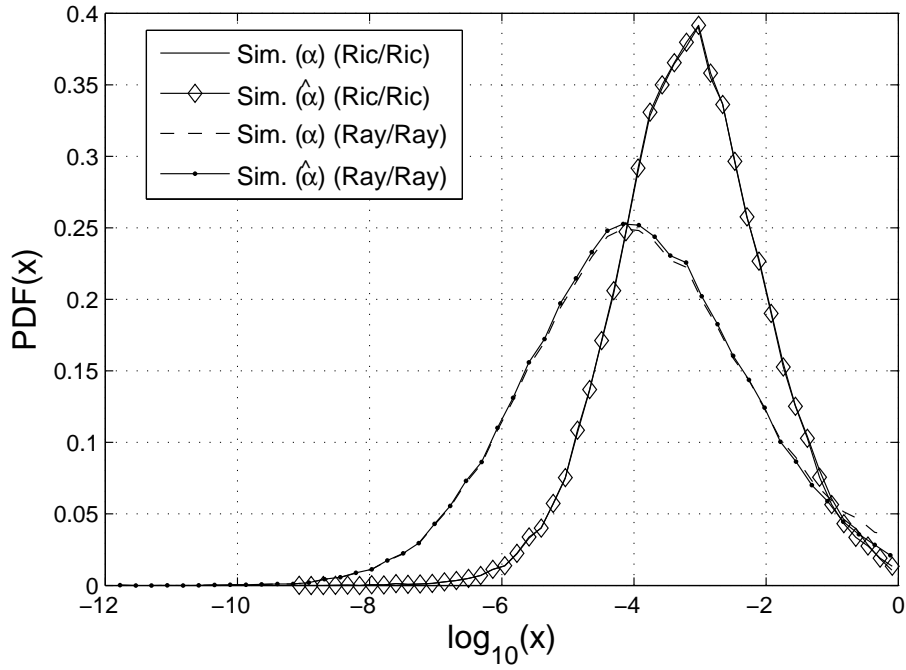}
\caption{PDFs of $\log_{10}(\alpha)$ and its approximation
$\log_{10}(\hat{\alpha})$. We use the default parameters for the
Ray/Ray curve, whereas, for Ric/Ric we have taken $\sigma = 4$ dB,
$\gamma = 2.5$ and $K = 5$ dB.} \label{fig3}
\end{figure}
\begin{figure}[t]
\centering
\includegraphics[width=0.75\columnwidth]{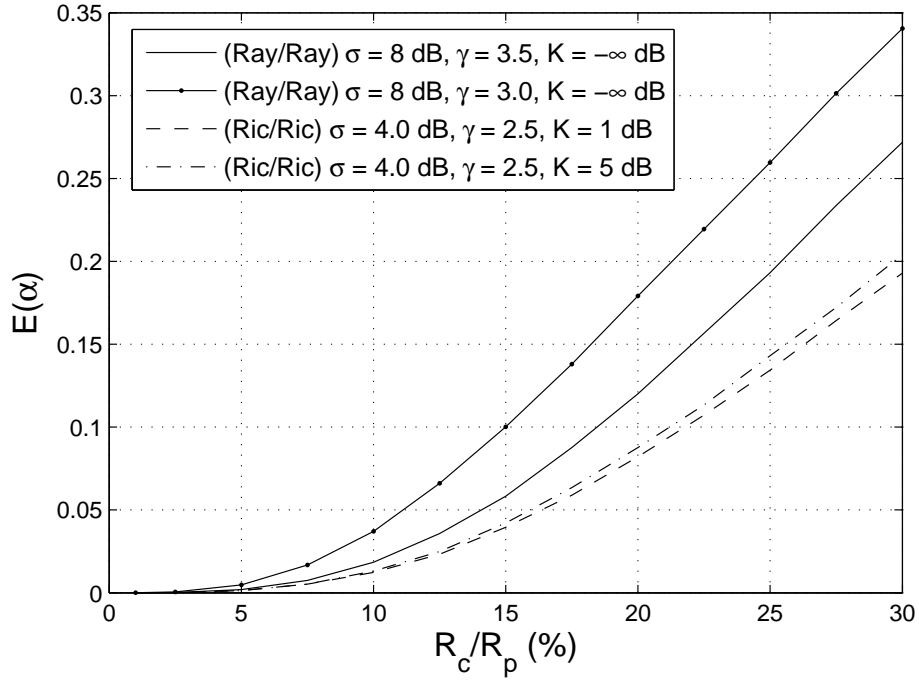}
\caption{Mean value of the power loss parameter, $\alpha$, as a
function of the ratio $\frac{R_c}{R_p}$ for different fading
scenarios.} \label{fig5}
\end{figure}
\begin{figure}[t]
\centering
\includegraphics[width=0.75\columnwidth]{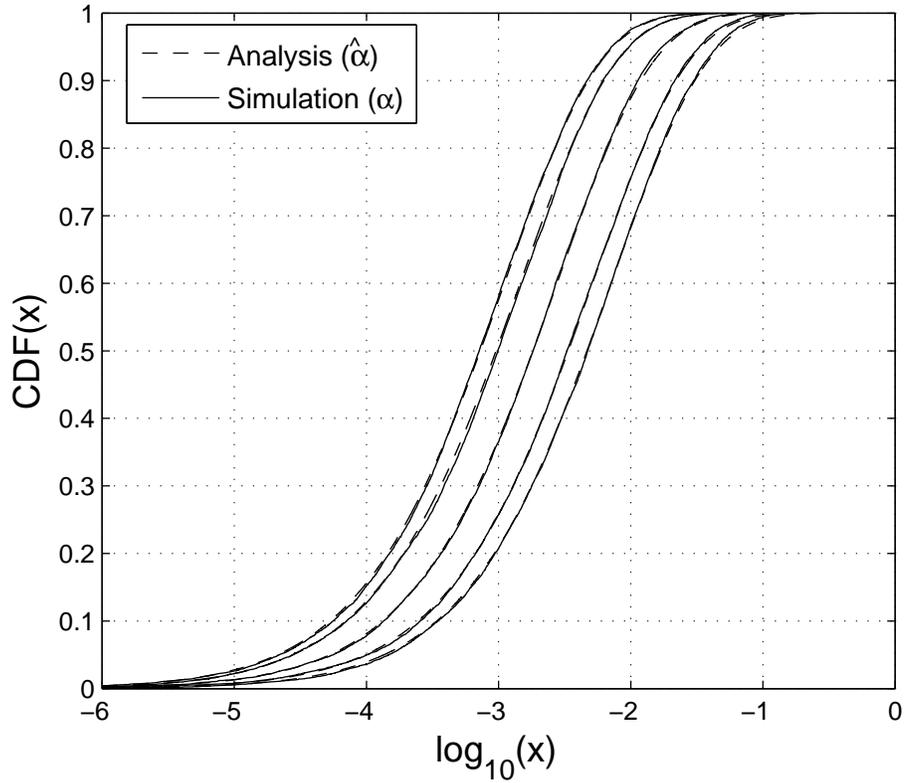}
\caption{Comparison of the exact and analytical CDFs of the power
loss factor on a logarithmic scale for fixed link gains. Results are
shown for 5 drops for the case of Ray/Ray fading.} \label{fig6}
\end{figure}
\begin{figure}[t]
\centering
\includegraphics[width=0.75\columnwidth]{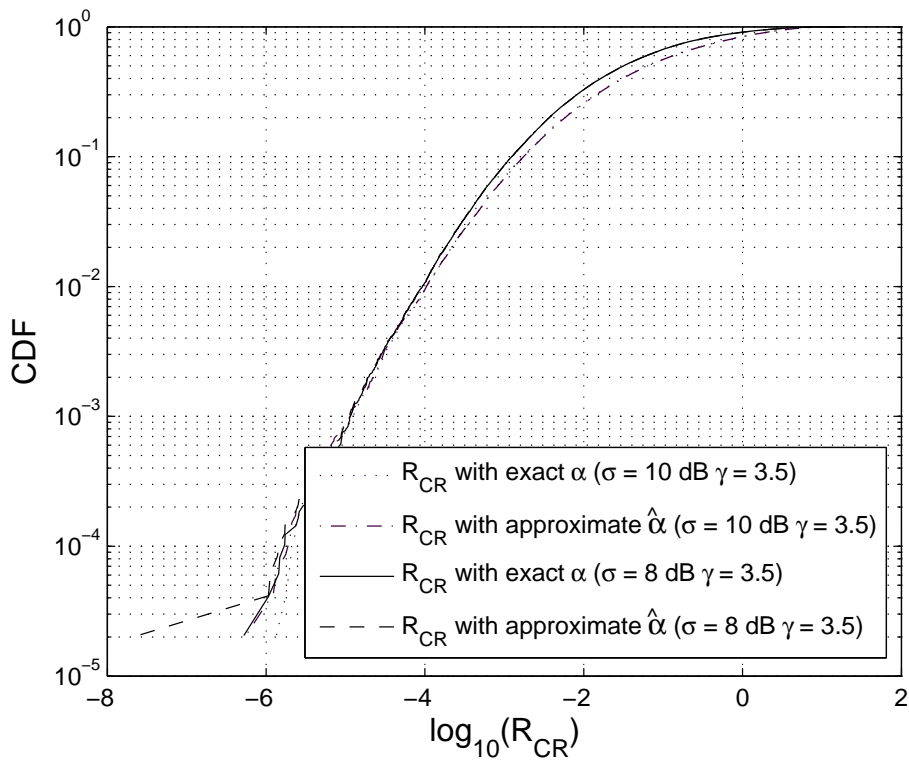}
\caption{CDF of the CR rates with the exact $\alpha$ and the
approximate $\hat{\alpha}$ for Ray/Ray fading.} \label{fig4}
\end{figure}
\begin{figure}[t]
\centering
\includegraphics[width=0.75\columnwidth]{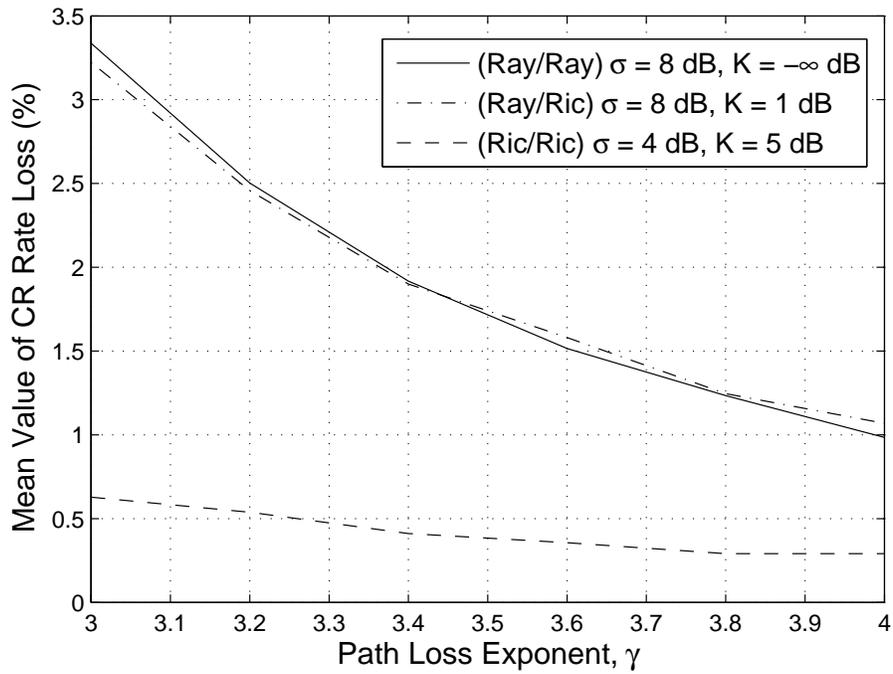}
\caption{Mean value of the CR rate loss as a function of $\gamma$
for different fading conditions.} \label{fig7}
\end{figure}
\begin{figure}[t]
\centering
\includegraphics[width=0.75\columnwidth]{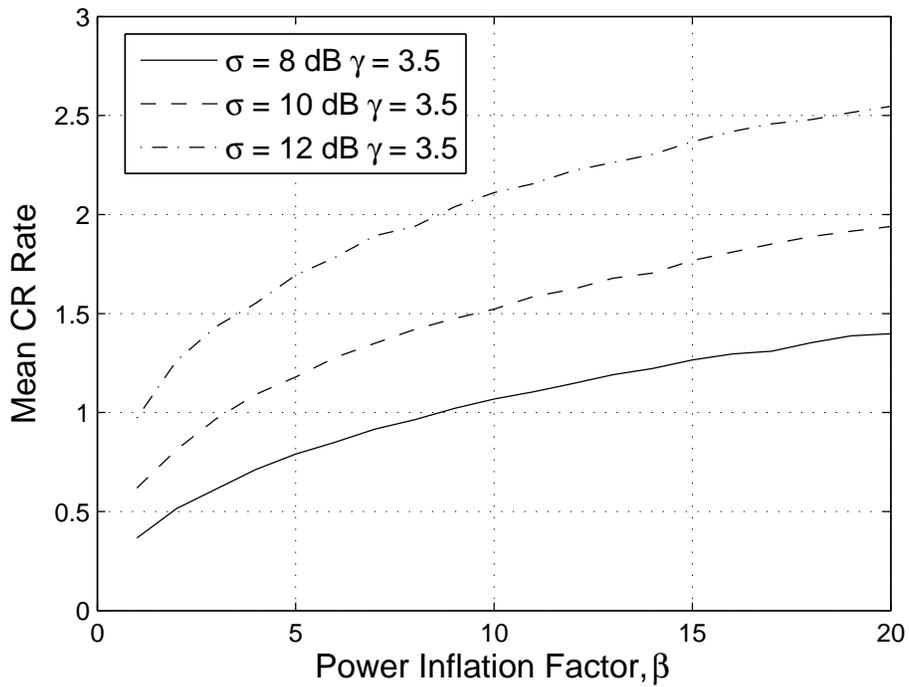}
\caption{Variation of the mean CR rate with the power inflation
factor, $\beta$ for Ray/Ray fading case.} \label{fig8}
\end{figure}
\end{document}